\documentstyle[12pt,epsfig]{article}
\input epsf

\newcommand{\be}[1]{
\begin{eqnarray}\label{#1}}
\newcommand{\ee}{\end{eqnarray}}
\newcommand{\ci}[1]{\cite{#1}}
\newcommand{\re}[1]{(\ref{#1})}

\newcommand\pbar{\bar{\psi}}
\newcommand\p{\psi }

\newcommand{\ba}{\begin{array}}
\newcommand{\ea}{\end{array}}

\def\Dirac#1{#1\hskip-6pt/}

\newcommand{\insertfig}[2]{\mbox{\epsfxsize=#1cm \epsfbox{#2.eps}}}

\begin{document}
\rightline{RUB-TPII-21/03}

\begin{center}
{\Large Soft Pion Emission from the Nucleon \\
Induced by Twist-2 Light-Cone Operators}\\[0.5cm]

 N.~Kivel$^{a,b}$, M.V.~Polyakov$^{a,c}$, S.~Stratmann$^{b}$ \\[0.3cm]

\footnotesize\it $^a$ Petersburg
Nuclear Physics Institute, Gatchina, St. Petersburg 188350,
Russia\\
\footnotesize\it $^b$ Institute for Theoretical Physics II, Ruhr
University Bochum, Germany
\\
\footnotesize\it $^c$
 Institut de Physique, B5a,
                  Universit\'e  de Li\`ege au Sart Tilman,
                  B-4000 Li\`ege 1 Belgium\\

\end{center}
\begin{abstract}
We compute the matrix elements for pion emission from the nucleon
induced by  twist-two light-cone operators in the kinematics where
the pion is close to threshold. It is shown that the results of
heavy baryon ChPT are in agreement with the results from the
soft-pion theorem approach. Therefore the amplitude of soft-pion
emission is directly related to the generalized parton distributions in
the nucleon.
\end{abstract}

\section{Introduction}

Hard processes are known to provide us with valuable information
about the quark and gluon structure of hadrons in terms of parton
distributions and distribution amplitudes. The generalized
parton distributions (GPDs) \cite{Mul94,Ji97b,Rad97,Col97}, entering
the QCD description of
hard exclusive processes, interpolate, in a sense,
between usual parton distributions, distribution amplitudes,
and elastic hadron form factors
(for a review see e.g.~\cite{revhard}).
GPDs are low energy observables and therefore their
dependence on the quark mass, small momentum transfer,
etc.~can be studied with  help
of the chiral perturbation theory (ChPT).

The main source of information about GPDs are exclusive
reactions which can be studied at many experimental
facilities \ci{exper}.
At the same time the existing experiments suffer from an
 ambiguity due to errors in the identification of the recoiling nucleon.
The missing mass technique used for this purpose cannot
distinguish between, for instance, one nucleon and a nucleon with
a pion produced close to the threshold. An estimate of such a
contamination was suggested in \ci{VanG} using the approach of
current algebra and  PCAC (= soft-pion theorems). Recently,
results of \ci{VanG} were criticized in \ci{Savage}. The authors
of \ci{Savage} used the chiral perturbation theory to calculate the
same matrix elements of twist-2 operators as in \ci{VanG}. It was
claimed  that the two results do not match each other. The aim of
our paper is to clarify this situation.

In our presentation we consider for simplicity only  twist-2
{\em isovector} operators, a generalization to other operators is trivial.
 We construct the corresponding effective operators in ChPT
in Section 2 and calculate their
matrix elements for soft pion emission from the nucleon.
In Section 3 we present soft-pion theorem (SPT) results
and perform a comparison with the ChPT calculations.

\subsection{GPDs in the chiral limit}

The nonlocal twist-2 light-cone matrix elements are parametrized in terms of the GPDs $H$, $E$, $\widetilde H$ and $\widetilde E$ as
\be{Fdef}
&&
(\bar P\cdot n)
\int^\infty_{-\infty}
\frac{d\lambda}{2\pi}e^{-ix\lambda(\bar P\cdot n)} \langle \bar P+\Delta/2|
\pbar(\lambda n/2)\Dirac{n}\frac{\tau^{3}}{2}\p
\left(-\lambda n/2\right)
 | \bar P-\Delta/2\rangle =
 \nonumber \\[4mm] &&
\bar N(\bar P+\Delta/2)
\biggr[
\Dirac{n} H(x,\xi,\Delta^2)+
\frac{i\sigma^{\mu\nu}{n_\mu\Delta_\nu}}{2M} E(x,\xi,\Delta^2)
\biggl]
\frac{\tau^{3}}{2}N(\bar P-\Delta/2)\, ,
\ee
\be{Ftlddef}
&&
(\bar P\cdot n)
\int^\infty_{-\infty}\frac{d\lambda}{2\pi}
e^{-ix\lambda(\bar P\cdot n)} \langle
 \bar P+\Delta/2| \pbar(\lambda n/2)\Dirac{n}\gamma_5
\frac{\tau^{3}}{2}\p (-\lambda n/2)
 | \bar P-\Delta/2\rangle =
\nonumber \\[4mm] &&
\bar N(\bar P+\Delta/2)
\biggl[
\Dirac{n}\gamma_5\widetilde H(x,\xi,\Delta^2)+
\gamma_5\frac{(\Delta\cdot n)}{2M}\widetilde E(x,\xi,\Delta^2)
\biggr]
\frac{\tau^{3}}{2}N(\bar P-\Delta/2)
\, ,
\ee
where $n$ is a light-cone vector, $n^2=0$, defining the separation
 of the fields
in the operators; $\bar N$ and $N$ are nucleon spinors;
 $\xi$ denotes the
standard skewedness variable $\xi=-\frac{1}{2}
{(\Delta \cdot n)}/{(\bar P\cdot n)}$. The variable
$x$ has the meaning of  momentum fraction carried by the quark with respect to
the average momentum $\bar P$ of the nucleon.

The following sum rules relate the GPDs to local form factors:
\be{lsrV}
&&\int_{-1}^{1}dx\, H(x,\xi,\Delta^2)=F^{p}_1(\Delta^2)-F^{n}_1(\Delta^2),
\\
&&\int_{-1}^{1}dx\, E(x,\xi,\Delta^2)=F^{p}_2(\Delta^2)-F^{n}_2(\Delta^2),
\\ &&
F^{p}_1(0)=1,\, F^{n}_1(0)=0,\quad
F^{p}_2(0)-F^{n}_2(0)=\kappa_p-\kappa_n\approx 3.706,
\ee
\be{lsrA}
&&\int_{-1}^{1}dx\, \widetilde H(x,\xi,\Delta^2)=G_A(\Delta^2),
\\ &&
 \int_{-1}^{1}dx\, \widetilde E(x,\xi,\Delta^2)=G_P(\Delta^2),
\\ &&
G_A(0)=g_A,\,\quad
G_P(\Delta^2)\approx \frac{4M^{2}g_A}{\Delta^2-m_\pi^{2}},
\ee
where we use the notations $M$ and $m_\pi$ for the
masses of the nucleon and
the pion, respectively.
To get a connection with heavy baryon ChPT we shall perform a
non-relativistic expansion of the non-local matrix elements.
For this purpose, we choose the Breit frame where
\be{Breit}
\bar P=[M+O(\Delta^2/M)]v,\quad v=(1,0,0,0),\quad (\Delta\cdot v)=0.
\ee
In this case, the  momentum transfer is  small ,
$\Delta\sim {\cal O}(\varepsilon)$, where as usual in ChPT $\varepsilon$ is a small
generic momentum. We shall also introduce the
standard large mass decomposition for the nucleon spinors:
\be{hns}
N(\bar P-\Delta/2)=\left(1-\frac{\Delta_\mu\gamma^\mu}{4M}\right)N_v+{\mathcal O}(\Delta^2),
\quad
\Dirac{v}N_v=N_v .
\ee
The power counting  for the light-cone variables is as follows:
\be{count}
\mbox{momentum fraction }&& x\sim {\cal O}(1),\\
\mbox{momentum transfer }&& \Delta \sim {\cal O}(\varepsilon),\\
\mbox{skewedness }&& \xi\sim {\cal O}(\Delta/M).
\ee
Performing an expansion of \re{Fdef} and \re{Ftlddef} with respect
to the
small parameter $\Delta$ we arrive at the non-relativistic expansion of the GPDs.
The  result can be written as
\be{Vch}
(\bar P\cdot n)
\lefteqn{\int^\infty_{-\infty}
\frac{d\lambda}{2\pi}e^{-ix\lambda(\bar P\cdot n)}
 \langle \bar P+\Delta/2|
\pbar(\lambda n/2)\Dirac{n}\frac{\tau^{3}}{2}\p (-\lambda n/2)
 |\bar P-\Delta/2\rangle} & &
\nonumber
\\ &=&
 q(x) (v\cdot n)\bar N_v \frac{\tau^{3}}{2} N_v+ {\cal O}(\Delta)\, ,
\\
(\bar P\cdot n)
\lefteqn{\int^\infty_{-\infty}\frac{d\lambda}{2\pi}e^{-ix\lambda(\bar P\cdot n)}
\langle
 \bar P+\Delta/2| \pbar(\lambda n/2)\Dirac{n}\gamma_5\frac{\tau^{3}}{2}\p
 (-\lambda n/2)
 | \bar P-\Delta/2
 \rangle}
&&
\nonumber
\\&=&
\Delta q(x)\bar N_v(S\cdot n)\tau^{3} N_v-
\delta(x)\frac{(\Delta\cdot n)g_A}{\Delta^2-m_\pi^{2}}
\bar N_v(S\cdot\Delta)\tau^{3} N_v
+ {\cal O}(\Delta)\, ,
\label{Ach}
\ee
where $q(x)=H(x,0,0)$ and $\Delta q(x)=\widetilde H(x,0,0)$ are the isovector  vector and axial-vector
forward distributions in the chiral limit and
 $S_\mu=\frac{i}{2}\gamma_5\sigma_{\mu\nu}v^\nu$ is the covariant spin matrix.

\section{Light-cone operators in the effective theory}

In this section we briefly discuss the form of the isovector operators in terms of effective fields.
We consider the $SU(2)$ flavor sector with the leading order effective Lagrangian
\be{action}
{\cal L}_{eff}=\frac{F_\pi^2}{4}{\rm Tr}
(\partial_\mu U\partial^\mu U^\dagger+
\chi U^\dagger+\chi^\dagger U)
\, + \bar H_v[(iv\cdot D)+g_A (u\cdot S)]H_v,
\ee
where as usual
\be{fields}
U(x)=\exp(i\pi^a(x) \tau^a/F_\pi),\quad \chi=2B\,{\rm diag}(m_u, m_d),
\quad U=u^2,\\
u_\mu=i[u^\dagger\partial_\mu u-u\partial_\mu u^\dagger],\quad
\Gamma_\mu=\frac{1}{2}[u^\dagger\partial_\mu u+u\partial_\mu u^\dagger].
\ee
 $D_\mu=\partial_\mu+\Gamma_\mu$ is the chiral covariant
 derivative and $H_v$ denotes the nucleon
field in the  heavy baryon approach.
Under the chiral $SU(2)_L\times SU(2)_R$ transformation the  fields change according to
\be{trsfm}
U'(x)=RUL^\dag,\quad H'_v=K(R,L,U)H_v,\quad
u'=Ru K^\dag=KuL^\dag,
\ee
where the transformation matrix $K$ depends not only on the group elements
$L$ and $R$ but also on the pion field $U$ \ci{WCWZ}.
For simplicity, we do not
consider contributions from spin-$\frac{3}{2}$
$\Delta$-resonances, since they are irrelevant for the subject of our discussion.

Our task is to find the leading-order expressions for the effective operators that correspond to the following light-cone QCD operators:
\be{Ops}
&&V^a(\lambda)=\pbar(\lambda n/2)\Dirac{n}\frac{\tau^{a}}{2}
\p(- \lambda n/2),
\\
&&A^a(\lambda)=\pbar( \lambda n/2)
\Dirac{n}\gamma_5\frac{\tau^{a}}{2}\p(-\lambda n/2).
\label{Ax}
\ee
We remark that we shall not perform the expansion in local operators
because the non-local notation is simpler.

The effective operators have to respect certain properties of their QCD counterparts.
First, we note that the QCD operators \re{Ops} and \re{Ax} are isovectors, and that they transform with respect to parity according to
\be{parity}
V^a(\lambda)\rightarrow V^a(\lambda),\qquad A^a(\lambda)\rightarrow -A^a(\lambda).
\ee
Taking these properties into account, we find that the effective operators can be constructed from appropriate building blocks in the following way:
\be{VGf}
&&
V^a(\lambda)= \frac{(v\cdot n)}{4}C_1(\lambda)\,
\bar H_v[ u^{\dag}\tau^a u +u\tau^a u^{\dag}]H_v+
\frac12 C_2(\lambda)\, \bar H_v(S\cdot n)[ u^{\dag}\tau^a u -u\tau^a u^{\dag}]H_v+
\nonumber \\ &&
\mskip 200mu \mbox{(pion part)},
\\[4mm] &&\label{AGf}
A^a(\lambda)=\frac{(v\cdot n)}{4}C_3(\lambda)\,
\bar H_v[ u^{\dag}\tau^a u -u\tau^a u^{\dag}]H_v+
 \frac12C_4(\lambda)\,\bar H_v(S\cdot n)[ u^{\dag}\tau^a u +u\tau^a u^{\dag}]H_v+
\nonumber \\ &&
\mskip 200mu \mbox{(pion part)},
\ee
where $C_i$, $i=1,2,3,4$, are unknown functions of the variable $\lambda$ and, for brevity,
we do not write explicitly pure pion contributions.

The number of functions can be reduced further if we consider that
under global $SU(2)_L \times SU(2)_R$ rotations the operator
combinations $V^a(\lambda)\pm A^a(\lambda)$ transform as pure right-
and left-handed quantities, respectively:
\be{sumR}
A^a(\lambda)+V^a(\lambda) & = &
\pbar_R(\lambda n/2)\Dirac{n}\tau^{a}\p_R(-\lambda n/2),\\
A^a(\lambda)-V^a(\lambda) & = &
 -\pbar_L(\lambda n/2)\Dirac{n}\tau^{a}\p_L(-\lambda n/2),
\ee
where we use the definitions $\psi_{R,L}=\frac{1\pm \gamma_5}{2}\psi$.
According to the transformation laws given in (\ref{trsfm}), the same
is true for certain combinations of the effective fields.
Namely, for an arbitrary $SU(2)_L \times SU(2)_R$
transformation it turns out that
\be{trsfmeff}
( \bar H_v u^\dag \tau^a u H_v)' & = & \bar H_v u^\dag R^\dag \tau^a R u H_v,\\
( \bar H_v u \tau^a u^\dag H_v)' & = & \bar H_v u L^\dag \tau^a L u^\dag H_v.
\ee
Hence the correct transformation behavior of
$V^a(\lambda)\pm A^a(\lambda)$ can be reproduced
only if the coefficient functions obey $C_1=C_3$ and $C_2=C_4$.

The remaining two functions are determined by comparing the nucleon
matrix elements of the operators in the effective theory with the
 Fourier transforms of the leading order expressions (\ref{Vch})
 and (\ref{Ach}), which read
\be{Vme}
&&\langle \bar P+\Delta/2|V^a(\lambda)|\bar P-\Delta/2\rangle=
 (v\cdot n)Q(\lambda)\bar N_v \frac{\tau^{a}}{2} N_v,
\\ &&
\langle \bar P+\Delta/2|A^b(\lambda)|\bar P-\Delta/2\rangle=
\widetilde Q(\lambda)\bar N_v(S\cdot n)\tau^{b} N_v-
\frac{(\Delta\cdot n)g_A}{\Delta^2-m^{2}}
\bar N_v(S\cdot \Delta)\tau^{b} N_v.
\label{Ame}
\ee
Here, the functions $Q$ and $\widetilde Q$ are
Fourier transforms of the cor\-res\-pon\-ding parton distributions:
\be{FT}
Q(\lambda)=\int_{-1}^{1} dx\, e^{ix\lambda(\bar P\cdot n)}q(x),\\
\widetilde{Q}(\lambda)=\int_{-1}^{1} dx\, e^{ix\lambda(\bar P\cdot n)}\Delta q(x).
\ee
Consequently, the functions $C_1$ and $C_2$ are fixed as
\be{fix}
C_1(\lambda)=Q(\lambda),\quad C_2(\lambda)=\widetilde Q(\lambda).
\ee
Therefore, the final leading order expressions for the effective operators are
\be{Vop}
V^a(\lambda)&=& \frac{1}{4}(v\cdot n)Q(\lambda)
\bar H_v[ u^{\dag}\tau^a u +u\tau^a u^{\dag}]H_v+
\frac12\widetilde Q(\lambda)
\bar H_v(S\cdot n)[ u^{\dag}\tau^a u -u\tau^a u^{\dag}]H_v+
\nonumber \\
&&
\phantom{\frac{1}{2}(v\cdot n)Q(\bar \lambda)
N_v[ u^{\dag}\tau^a u +u\tau^a u^{\dag}]N_v}
\mbox{(pion term)},
\ee
\be{Aop}
A^a(\lambda)&=&\frac{1}{4}(v\cdot n)Q(\lambda)
\bar H_v[ u^{\dag}\tau^a u -u\tau^a u^{\dag}]H_v+
\frac12\widetilde Q(\lambda)\bar
H_v(S\cdot n)[ u^{\dag}\tau^a u +u\tau^a u^{\dag}]H_v+
\nonumber \\ && \phantom{\frac{1}{2}(v\cdot n)Q(\lambda) N_v[
u^{\dag}\tau^a u +u\tau^a u^{\dag}]N_v} \mbox{(pion term)}. \ee

 In conclusion, let us emphasize that
the number of unknown functions in \re{VGf} and \re{AGf}
was fixed using the behavior of the operators $V^a(\lambda)$ and $A^a(\lambda)$ under chiral transformations.
In the paper \ci{Savage} this was not done and as a result it was claimed that the
final expressions for the effective operators contain additional unknown functions.
Hence, the conclusion
made in \ci{Savage} about
contradictions of the chiral perturbation theory with
the soft-pion theorem derived in \ci{VanG}
seems to be incorrect and
has to be reconsidered. Below we shall show that at tree level the correct effective operators
give the same answer as the soft-pion theorem, as  usually expected.

The explicit form of the pion terms in \re{Vop} and \re{Aop} is not
important. In the kinematics of the heavy baryon approach we shall need
for actual calculations only contributions arising
from the vertex corresponding to local vector and axial-vector currents.
Therefore, here we skip the discussion of the
structure of such terms. Let us only mention that similar issues in
the pure
pion sector were discussed in \ci{KM}. In particular, it
was demonstrated
that the chiral perturbation theory and SPTs, derived from
the current algebra approach, give identical results.

\section{Soft pion emission}

In this section we evaluate  matrix elements of the
effective operators \re{Vop} and \re{Aop} between an
initial nucleon state and a final nucleon-pion state.
Relevant diagrams are given in Fig.1.
\begin{figure}[t]
\unitlength1mm
\begin{center}
\hspace{0cm} \insertfig{8}{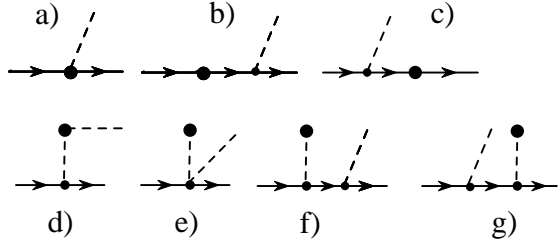}
\end{center}
\vspace{-0.5cm} \caption[dummy]{\small
Chiral perturbation theory diagrams. The large  black blobs denote the
vertices of the effective operator. Dashed lines and solid lines with arrows represent the pion and nucleon, respectively.
\label{diag}}
\end{figure}
A direct calculation of these diagrams gives
\be{Vres}
&&
 \langle \bar P+\Delta/2,\pi^a(k)|
V^3(\lambda)
 | \bar P-\Delta/2\rangle =
 -\frac{1}{F_\pi}\widetilde Q(\lambda) \epsilon_{3ab}\bar
 N_v(S\cdot n) \tau^{b} N_v+
 \nonumber \\[4mm] &&
\frac{g_A}{F_\pi}\frac{(v\cdot n) Q(\lambda) }{(k\cdot v)}
 \epsilon_{3ab}\bar N_v(S\cdot k) \tau^{b}N_v-
\frac{g_A}{F_\pi}\frac{(k-\Delta)\cdot n}{\Delta^2-m_\pi^2}
\epsilon_{3ab}N_v(S\cdot\Delta) \tau^{b}N_v,
\ee
where $k$ denotes the small pion momentum and all other notations are
 the same as before. The first term on the right-hand side of
 \re{Vres} corresponds to
the contribution of diagram a) in Fig.1, the second term is
the result of the
sum of  contributions  b) and c),
 and the third term originates from  diagram d).
 Diagrams e), f), and g) do not contribute to the matrix element of
 the vector operator.

For the axial-vector isovector operator we obtain
\be{Ares}
&&
 \langle \bar P+\Delta/2,\pi^a(k)|
A^3(\lambda)
 | \bar P-\Delta/2\rangle =
 -\frac{1}{2F_\pi} (v\cdot n) Q(\lambda) \epsilon_{3ab}\bar
 N_v \tau^{b} N_v+
 \nonumber \\[4mm] &&
\frac{ig_A}{F_\pi}\frac{\widetilde Q(\lambda)}{(k\cdot v)}
 \bar N_v[(S\cdot k)(S\cdot n)\tau^a \tau^{3}-(S\cdot n)(S\cdot k)
 \tau^3 \tau^{a}]N_v+
\nonumber \\[4mm] &&
\frac{ig_A^2}{F_\pi}\frac{1}{(k\cdot v)}
\frac{(k+\Delta)\cdot n}{(k+\Delta)^2-m_\pi^2}
\bar N_v[(S\cdot(k+\Delta))(S\cdot k)\tau^3 \tau^{a}-
(S\cdot k)(S\cdot(k+\Delta))\tau^a \tau^{3}]N_v+
\nonumber \\[4mm] &&
\frac{1}{2F_\pi}\frac{(k+\Delta)\cdot n\,(k\cdot v)}{(k+\Delta)^2-m_\pi^2}
\epsilon_{3ab}\bar N_v \tau^{b}N_v.
\ee
Again, the first term on the right hand side of \re{Ares} is due to a),
the second is due to b)+c). Further, the third term comes from f)+g)
and the fourth
is the contribution from e).
In the local limit we have $Q(0)=1$ and $\widetilde Q(0)=g_A$, so
 formulas \re{Vres} and \re{Ares}
 reproduce the correct expressions for the local currents
 which satisfy the following properties:
\be{curcon}
\partial_\mu \pbar\gamma_\mu\frac{\tau^{3}}{2}\p=0,\quad
\partial_\mu \pbar\gamma_\mu\gamma_5\frac{\tau^{3}}{2}\p=O(m_\pi^2).
\ee

Our goal is to compare the ChPT results \re{Vres} and
\re{Ares} with the corresponding soft-pion theorems (SPTs).
 A detailed derivation of the  SPT formulas
will be given in a separate publication \ci{PoSt}.
In this presentation we shall
only discuss the connection of the SPT approach with ChPT.

Before the presentation of the SPT formulas we
introduce a set of slightly different notations which are convenient
in this case. We shall use

\begin{equation}
p'=\bar P+\Delta/2,\quad p=\bar P-\Delta/2,\quad p'-p=\Delta,
\end{equation}
\begin{equation}
Q^a_5=\int d^3x\, \pbar(x)\,\gamma_0\gamma_5\frac{\tau^a}{2}\p(x),
\end{equation}
\be{new}
&&\langle p'|V^3(\lambda)| p\rangle=
\bar N(p')\,\Gamma^3_V(p',p,\lambda)\,N(p),
\\ &&
\Gamma^3_V(p',p,\lambda)\equiv
\int_{-1}^1 dx\, e^{ix\lambda(\bar P\cdot n)}
\biggr[
\Dirac{n} H(x,\xi,\Delta^2)+
\frac{i\sigma^{\mu\nu}{n_\mu\Delta_\nu}}{2M} E(x,\xi,\Delta^2)
\biggl]
\frac{\tau^{3}}{2},
\\[5mm] &&
\langle p'|A^3(\lambda)| p\rangle=
\bar N(p')\,\Gamma^3_A(p',p,\lambda)\,N(p),\\ &&
\Gamma^3_A(p',p,\lambda)\equiv
\int_{-1}^1 dx\, e^{ix\lambda(\bar P\cdot n)}
\biggl[
\Dirac{n}\gamma_5\widetilde H(x,\xi,\Delta^2)+
\Dirac{\Delta}\gamma_5\frac{(\Delta\cdot n)}{(2M)^2}\widetilde E(x,\xi,\Delta^2)
\biggr]
\frac{\tau^{3}}{2}.
\ee
With these definitions the soft-pion theorems for the matrix elements of the isovector operators read
\be{sptV}
\lefteqn{\langle p',\pi^a(k)|V^3(\lambda)| p\rangle =
-\frac{i}{F_\pi}\langle p'|[Q^a_5,V^3(\lambda)]| p\rangle } & &
\nonumber \\
&&
+
\frac{g_A}{2F_\pi}\bar N(p')
\left[
  \Dirac{k}\gamma_5\tau^a
  \frac{i(\Dirac{p}'+M)}{2(p'\cdot k)}
  \Gamma_V^3(p'+k,p,\lambda)
 + \Gamma_V^3(p',p-k,\lambda)\,
  \frac{i(\Dirac{p}+M)}{-2(p\cdot k)}\Dirac{k}\gamma_5\tau^a
\right]
N(p) \nonumber \\
&&
+
\frac{g_A}{2F_\pi}
\bar N (p')\Dirac{\Delta}\gamma_5\tau^b N(p)
\frac{i}{\Delta^2-m_\pi^2}
i(n\cdot k) \epsilon_{3ab}
+{\cal O}(\varepsilon),
\ee

\be{sptA}
\lefteqn{\langle p',\pi^a(k)|A^3(\lambda)| p\rangle =
 -\frac{i}{F_\pi}\langle p'|
 [Q^a_5,A^3(\lambda)]| p\rangle} & & \nonumber \\
&&
+
\frac{g_A}{2F_\pi}\bar N(p')
\left[\Dirac{k}\gamma_5\tau^a
 \frac{i(\Dirac{p}'+M)}
 {2(p'\cdot k)}\Gamma_A^3(p'+k,p,\lambda)
 +\Gamma_A^3(p',p-k,\lambda)
 \frac{i(\Dirac{p}+M)}{-2(p\cdot k)}\Dirac{k}
 \gamma_5\tau^a
\right] N(p) \nonumber \\
&&
-
\frac{\epsilon_{3ab}}{4F^2_\pi}
\bar N(p')(\Dirac{\Delta}+2\Dirac{k})\tau^b N(p)
\frac{i}{(k+\Delta)^2-m_\pi^2}
i F_\pi (n\cdot(\Delta +k))
+{\cal O}(\varepsilon).
\ee
In the derivation of these formulas it is essential to assume
\begin{equation}
m_\pi = {\cal O} (\varepsilon), \qquad k={\cal O}(\varepsilon),
\end{equation}
while in contrast to the ChPT counting \re{count} here it is not necessary to demand $\Delta^2={\cal O}(\varepsilon^2$).
Note that in the particular range where $|\Delta^2|\gg \varepsilon^2$, the explicit
pion pole terms in (\ref{sptV}) and (\ref{sptA}) are of order $k$, so within the overall accuracy of the SPTs they can be neglected.
Therefore, in this kinematical region, our results are in agreement with corresponding formulas in \ci{VanG}.

For a comparison with the ChPT results \re{Vres} and \re{Ares} we have to consider the region $\Delta^2={\cal O}(\varepsilon^2)$, i.e.~we have to perform a non-relativistic expansion of the SPTs \re{sptV} and \re{sptA}.
Let us show the procedure in some detail for the vector current formula
 \re{sptV}.
First, we have to evaluate the commutator $[Q^a_5,V^3(\lambda)]$.
Since we are only interested in the leading-order contribution, we
can restrict ourselves to the chiral limit where $Q^a_5$ is
time-independent.
Then the commutation relations of the fields,
\be{comf}
[Q^a_5,\p]=-\frac{\tau^a}2\gamma_5\p,\quad
[Q^a_5,\pbar]=-\pbar\gamma_5\frac{\tau^a}2,
\ee
can be used to calculate the commutator of $Q^a_5$ with the light-cone operator $V^3(\lambda)$. In this way we obtain for the first term on the right-hand side of \re{sptV}
\be{cterm1}
\lefteqn{-\frac{i}{F_\pi}\langle p'|[Q^a_5,V^3(\lambda)]| p\rangle =-\frac{\epsilon_{3ab}}{F_\pi}\langle p'|A^b(\lambda)| p\rangle
}&& \\
&&
=-\frac{\epsilon_{3ab}}{F_\pi}
\widetilde Q(\lambda)\bar N_v(S\cdot n)\tau^b N_v
+\frac{\epsilon_{3ab}}{F_\pi}\frac{(\Delta\cdot n)g_A}{\Delta^2-m_\pi^{2}}
\bar N_v(S\cdot \Delta)\tau^{b} N_v
+ {\cal O}(\varepsilon)\, ,
\label{c1end}
\ee
where we used \re{Ame} to pass from \re{cterm1} to \re{c1end}.
The first term of this contribution exactly equals that one of diagram a) in \re{Vres}.
The second term in combination with the pion pole in the last line of \re{sptV} is identical to diagram d).

Consider now the second line on the right hand side of \re{sptV}, which represents the contributions of the nucleon poles.
In the non-relativistic limit we obtain
\be{exp:npt}
&&
\frac{g_A}{2F_\pi}\bar N(p')
\left[\Dirac{k}\gamma_5\tau^a
 \frac{i(\Dirac{p}'+M)}{2(p'\cdot k)}
 \Gamma_V^3(p'+k,p,\lambda)
 +\Gamma_V^3(p',p-k,\lambda)
 \frac{i(\Dirac{p}+M)}{-2(p\cdot k)}\Dirac{k}\gamma_5\tau^a\right] N(p)
\nonumber \\
&=&
\frac{g_A}{F_\pi}\frac{(v\cdot n) Q(\lambda) }{(k\cdot v)}
 \epsilon_{3ab}\bar N_v(S\cdot k) \tau^{b}N_v+{\cal O}(\varepsilon).
\ee
Comparing with \re{Vres} we find that the nucleon pole terms
\re{exp:npt} correspond
to the contributions of diagrams b) and c) in Fig.1.


An analogous analysis can be performed for the axial-vector current as well.
Again, the final expression is
in agreement with the corresponding ChPT calculation \re{Ares}.
Therefore, we have demonstrated the agreement between results which have been obtained
by two technically different ways. We would like to stress that
this agreement is expected, because historically speaking, soft-pion theorems represented the first step in the development of ChPT.

\section{Conclusions}
Using the approach of chiral perturbation theory
we  calculated the soft pion emission from  nucleon induced by
 twist-2 light-cone operators.
It was shown that the  soft-pion theorem results are
in agreement with ChPT as it should be. We have found that
the contradictions that were reported in paper \ci{Savage}
originate from the incorrect representation of the effective twist-2
operators given in that paper.

\section{Acknowledgments}
We are thankful to M.~Savage for
 the useful correspondence and V.~Guzey for the comments about the text.
  M.P. is grateful to Prof.~K. Goe\-ke for
the warm hospitality in the Bochum University. Research
reported in this paper has been supported by the Kovalevskaja Program of
the Alexander von Humboldt Foundation, and by the
COSY-Julich project.

\end{document}